\documentclass[10pt, a4paper]{article}
\usepackage[utf8]{inputenc}
\usepackage{authblk}
\usepackage{cite}
\usepackage{fullpage}
\usepackage{amsmath}
\usepackage{graphicx}
\usepackage{tabularx}
\usepackage{titlesec}

\title{Predictive coding underlies adaptation in the \\subcortical sensory pathway} 
\author[12]{Alejandro Tabas}
\author[2]{Glad Mihai}
\author[1]{Stefan Kiebel}
\author[2]{Robert Trampel}
\author[1]{Katharina von Kriegstein}
\affil[1]{Department of Psychology, Technische Universit\"{a}t Dresden, Dresden, Germany}
\affil[2]{Max Planck Institute for Human Cognitive and Brain Sciences, Leipzig, Germany}
\date{}
\titleformat*{\section}{\large\bfseries}
\titleformat*{\subsection}{\bfseries}

\begin{document}

  \maketitle

	\abstract{
		The subcortical sensory pathways are the fundamental channels for mapping the outside world to our minds. Sensory pathways efficiently transmit information by adapting neural responses to the local statistics of the sensory input. The longstanding mechanistic explanation for this adaptive behaviour is that neuronal habituation scales activity to the local statistics of the stimuli. An alternative account is that neural coding is directly driven by expectations of the sensory input. Here we used abstract rules to manipulate expectations independently of local stimulus statistics. The ultra-high-field functional-MRI data show that expectations, and not habituation, are the main driver of the response amplitude to tones in the human auditory pathway. These results provide first unambiguous evidence of predictive coding and abstract processing in a subcortical sensory pathway, indicating that the brain only holds subjective representations of the outside world even at initial points of the processing hierarchy.
	}

  \section*{Introduction}

    Expectations have measurable effects on human perception; for instance, when disambiguating ambivalent stimuli like an object in the dark or spoken sentences in a noisy pub \cite{deLange2018}. The predictive coding theoretical framework \cite{Rao1999, Friston2005} formalises the active role of expectations on perception by suggesting that sensory neurons constantly match the incoming stimuli against an internal prediction derived from a generative model of the sensory input. This strategy increases the efficiency of encoding and naturally boosts the salience of unexpected events that often have strong relevance for behaviour and survival. Although predictive coding is has been shown for sensory processing in the cerebral cortex (see~\cite{Kok2015} for a review), the role of predictability in subcortical sensory coding is unclear \cite{Malmierca2019, Carbajal2018, Malmierca2015}. If coding at the subcortical pathway was based on expectations in the incoming stimuli, that would mean that the brain does not hold a veridical representation of the environment even at the very early points of the processing hierarchy. 

    Several studies in non-human mammals \cite{Ayala2015, Gao2014, Perez2012, Zhao2011, Bauerle2011, Antunes2010} as well as in humans \cite{Cacciaglia2015, Cornella2015, Escera2014, Grimm2011} have shown that single neurons and neuronal ensembles of subcortical sensory pathway nuclei exhibit stimulus specific adaptation (SSA). Neurons and neural populations showing SSA adapt to so called standards (frequently occurring stimuli) yet show restored responses to so-called deviants (rarely occurring stimuli) \cite{Ulanovsky2003, Antunes2010, Zhao2011}. In the auditory modality, SSA is typically elicited using sequences consisting of repetitions of a standard sound (typically a pure tone of a given frequency) incorporating a single, randomly located, deviant (a pure tone of the same duration and loudness but with a different frequency).

    Although SSA is often taken to support the view of predictive coding \cite{Carbajal2018, Malmierca2015, Cacciaglia2015}, it can also be explained in terms of stimulus-induced neural habituation \cite{Malmierca2014}, where neurons show decreased responsiveness to repeated stimuli independently of their predictability (see~\cite{GrillSpector2006, Kok2015} for reviews). Neural habituation has been proposed to be caused by synaptic fatigue \cite{Wang2014}, network habituation effects \cite{Eytan2003, Mill2011}, or sharpening of the receptive fields after stimulus repetition \cite{GrillSpector2006}. Similar adaptation dynamics to the local statistics of stimuli occurs even at the level of the retina \cite{Hosoya2005} and the cochlea \cite{Yates1990}. In contrast, the predictive coding framework \cite{Rao1999, Friston2005} suggests that neural activity represents prediction error and that such prediction error is minimal for predictable stimuli independently of their local statistics \cite{Malmierca2015}. 

    Distinguishing between these two mechanisms requires to manipulate predictability orthogonally to stimulus repetition and statistics \cite{Summerfield2008}. One way to do this is to control for behavioural expectations using abstract rules, an unresolved technical challenge for previous studies that mostly considered SSA in (often anaesthetised) animal models. Here, we used a novel paradigm (Figure~\ref{fig:design}A) in combination with ultra-high-field functional Magnetic Resonance Imaging (fMRI) in human subjects to disassociate the habituation and predictive coding views of adaptation in the auditory subcortical sensory pathway. We focused on the nuclei of the thalamus (medial geniculate body, MGB) and midbrain (inferior colliculus, IC) as they are the key nuclei of the ascending subcortical pathway that can be reliably investigated in human participants in vivo \cite{Sitek2019}.

  \section*{Experimental Design and Hypotheses}

\begin{figure}[h]
      \centering
      \includegraphics[width=0.5\columnwidth]{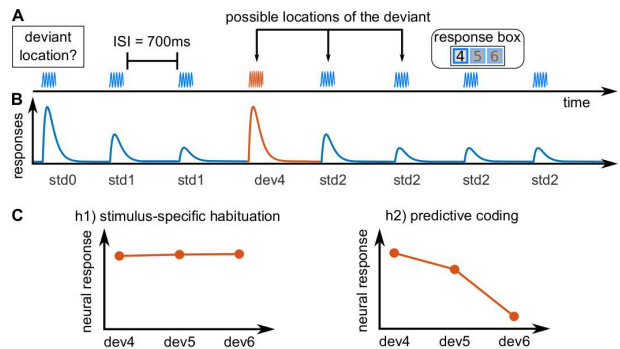}
      \caption{\textbf{Experimental design and hypotheses}. A) Example of a trial, consisting of a sequence of seven pure tones of a standard frequency (blue waveform) and one pure tone of a deviant frequency (red waveform), that could be located in positions 4, 5, or 6. Subjects had to report, in each trial, the position of the deviant. Each subject completed 240 trials in total, 80 per deviant location. All tones had a duration of 50\,ms and were separated by 700\,ms inter-stimulus-intervals (ISIs). B) Schematic view of the expected underlying responses in the auditory pathway for the sequence shown in A, together with the definition of the experimental variables ($std0$: first standard; $std1$: repeated standards preceding the deviant; $std2$: standards following the deviant; $dev\,x$: deviant in position $x$). C) Expected responses in the auditory pathway nuclei corresponding to the habituation (h1) and predictive coding (h2) hypotheses.
      \label{fig:design}}
    \end{figure}

    We measured blood-oxygenated-level-dependent (BOLD) responses in the human subcortical auditory pathway using 7 Tesla fMRI with a spatial resolution of 1.5\,mm isotropic. We recorded a slab comprising the MGB and the IC. Nineteen subjects listened to sequences of 8 pure tones (7 repetitions of a standard and one deviant tone; see Figure~\ref{fig:design}A). Subjects reported the position of the deviant for each sequence by pressing one button of a response box as quickly as possible. Expectations for each of the deviant positions were manipulated by two abstract rules that were disclosed to the subjects by stating: 1) all sequences have a deviant, and 2) the deviant is always located in position 4, 5, or 6. Note that, although the three deviant positions were equally likely at the beginning of the sequence, due to the two abstract rules the probability of finding a deviant in position 4 after hearing 3 standards is $1/3$, the probability of finding a deviant in position 5 after hearing 4 standards is $1/2$, and the probability of finding a deviant in position 6 after hearing 5 standards is $1$. Critically, habituation and predictive coding make opposing predictions for the responses at the different deviant positions (Figure~\ref{fig:design}C). According to the habituation hypothesis, deviants will elicit roughly similar responses independently of their position (or stronger responses for later deviants, see Supplementary Text S1). Conversely, under the predictive coding view the response is hypothesised to scale with the probability of finding a deviant in the target position, rendering responses to earlier deviants stronger in contrast to the later deviants. 

  \section*{Behavioural Responses}

    All subjects showed ceiling performances to all deviant positions ($97\pm3$\%, $98\pm3$\%, and $99\pm2$\% mean accuracies $\pm$ standard error of the mean, for deviants in positions 4, 5 and 6, respectively), indicating that subjects were attentive. Reaction times ($RT = 530\pm120\,$ms, $RT = 430\pm90\,$ms, $RT = 150\pm12\,$ms for deviants at positions 4, 5 and 6, respectively) were shorter for the more expected deviants, indicating a behavioural benefit of predictability. RTs were significantly shorter for deviants at positions 6 than for deviant at positions 4 and 5 (Cohen's $d = -1.02$ and $d = -1.00$, respectively; $p < 10^{-4}$ according to a Ranksum test with $N = 19$ samples, Bonferroni corrected for 3 comparisons). The RT difference between deviants 4 and 5 did not reach significance (p = 0.1, uncorrected; same test as above, Cohen's $d = 0.22$).

  \section*{Stimulus specific adaptation (SSA) in IC and MGB}

    We estimated BOLD responses to the different stimuli using a general linear model (GLM) with 6 different conditions: the first standard ($std0$), the standards after the first standard but before the deviant ($std1$), the standards after the deviant ($std2$), and deviants at positions 4, 5, and 6 ($dev4$, $dev5$, and $dev6$, respectively; Figure 1B). The conditions $std1$ and $std2$ were parametrically modulated according to their positions to account for possible variations in the responses over subsequent repetitions (see Methods; Figure~S1). 

\begin{figure}[h]
      \centering
      \includegraphics[width=0.75\columnwidth]{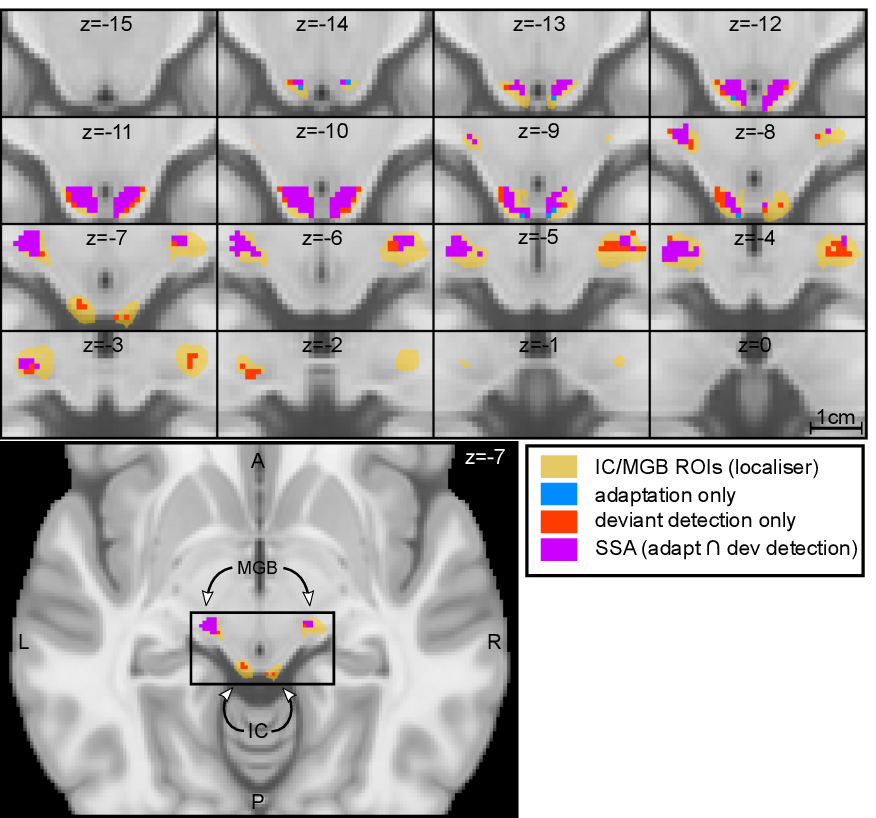}
      \caption{\textbf{Mesoscopic stimulus specific adaptation (SSA) in bilateral IC and MGB}. Regions within the MGB and IC ROIs showed adaptation to the repeated standards (adaptation; blue+purple) and deviant detection (red+purple). Stimulus specific adaptation (i.e., recovered responses to a deviant in voxels showing adaptation) occurred in bilateral MGB and IC (purple). Responses were computed by thresholding the contrast t-maps at \mbox{$\alpha = 0.05$}, Bonferroni corrected for the number of voxels in each of the anatomical ROIs.
      \label{fig:ROIs}}
    \end{figure}

    In the first step of the analysis, we determined those voxels within the ICs and MGBs that showed SSA at the mesoscopic level; i.e., that adapted to repeated stimuli and had restored responses to a deviant. We first identified the bilateral IC and MGB (IC and MGB ROIs; yellow patches in Figure~\ref{fig:ROIs}) based on structural MRI data and an independent functional localiser (see Methods). Within these ROIs, we tested: 1) for voxels with adapting responses to repeated standards (contrast $std0 > std2$); and 2) for voxels showing deviant detection, where the deviant elicited a stronger response than the repeated standards (contrast $dev4 > std2$). We found significantly adapting ($p < 0.0025$) and deviant detecting ($p < 0.00001$) voxels in all four anatomical ROIs (Table~\ref{tab:clusters}). We defined the set of voxels showing SSA as the intersection between the set of adapting and the set of deviant detecting voxels (purple in Figure~\ref{fig:ROIs}).

    \begin{table*}[b]
      \begin{tabularx}{\columnwidth}{rcccccc}
          contrast             &    ROI    & cluster size & MNI coordinates (mm) &  peak-level $p$-value  \\
          \hline
          adaptation           & left IC   &   88 voxels  &   $[ -3, -35, -10]$  & $p = 7.7\times10^{-6}$ \\
                               & right IC  &   74 voxels  &   $[  3, -35, -11]$  & $p = 4.8\times10^{-5}$ \\
                               & left MGB  &   85 voxels  &   $[-17, -24,  -5]$  & $p = 2.6\times10^{-5}$ \\
                               & right MGB &   21 voxels  &   $[ 16, -22,  -6]$  & $p = 2.4\times10^{-3}$ \\
          \\
          \hline
          deviant detection    & left IC   &  105 voxels  &   $[ -7, -33, -10]$  & $p = 2.2\times10^{-5}$ \\
                               & right IC  &   84 voxels  &   $[  4, -34, -11]$  & $p = 1.1\times10^{-6}$ \\
                               & left MGB  &   99 voxels  &   $[-13, -22,  -8]$  & $p = 6.0\times10^{-5}$ \\
                               & right MGB &   53 voxels  &   $[ 17, -22,  -6]$  & $p = 4.5\times10^{-6}$ \\
          \hline
        \end{tabularx}
      \caption{\textbf{Statistics and MNI coordinates of the adaptation and deviant detection contrasts in the four regions of interest.} All $p$-values Bonferroni corrected for the number of voxels in each anatomical ROI and further corrected for 8 comparisons. 
      \label{tab:clusters}}
    \end{table*}

  \section*{Adjudicating between habituation and predictive coding}

\begin{figure*}[t]
      \centering
      \includegraphics[width=\columnwidth]{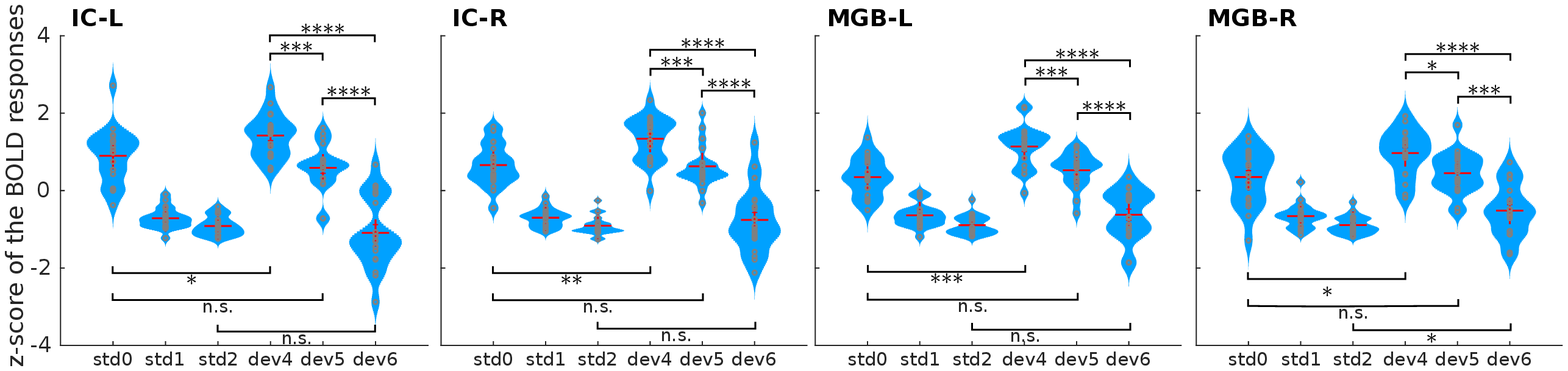}
      \caption{\textbf{BOLD responses in the four subcortical ROIs to the three different positions of the deviants.} Kernel density estimations of the distribution of $z$-scores of the estimated BOLD responses, averaged over voxels of each ROI, to the three deviant positions (dev4, dev5, dev6) in each four different ROIs: left and right IC, and left and right MGB (IC-L, IC-R, MGB-L, MGB-R). Responses to the three different standards ($std0$, $std1$, $std2$) are displayed for reference. Each distribution holds 19 samples, one per subject. Red crosses signal the mean and standard error of the distributions. * $p < 0.05$, ** $p <0.005$, *** $p < 0.0005$, **** $p < 0.00005$; all p-values are Holm-Bonferroni corrected for $8\times4=32$ comparisons. $Std0$, first standard; $std1$: standard(s) preceding the deviant; $std2$: standard(s) following the deviant; $dev4$, $dev5$ and $dev6$: deviants at positions 4, 5, and 6, respectively.
      \label{fig:betas}}
    \end{figure*}

    To test the two opposing hypotheses (Figure~\ref{fig:design}C), we studied the estimated BOLD responses to the different deviant positions in each SSA ROI of the ICs and MGBs (Figure~\ref{fig:betas}). On visual inspection, the response profile showed reduced responses for expected deviants, fitting with h2 (the predictive coding hypothesis; Figure~\ref{fig:design}C). Formal (Ranksum) statistical tests revealed significant differences in responses to the different deviant positions at \mbox{$\alpha = 0.05$} for all contrasts ($dev4 \leftrightarrow dev5$, $dev5 \leftrightarrow dev6$, $dev4 \leftrightarrow dev6$) in the four ROIs (\mbox{$p < 0.005$} (corrected for multiple comparisons), \mbox{$|d| > 1.00$}; for exact $p$-values and effect sizes see Table~\ref{tab:effectsize}). The results of these tests reject the habituation hypothesis, showing that MGB and IC mesoscopic responses to deviant tones cannot be explained by adaptation mechanisms only.

    \begin{table*}
      \begin{tabularx}{\columnwidth}{rcc|cc|cc}
               &                                        \multicolumn{6}{c}{\textbf{left IC}}                                        \\
          \hline
               &       \multicolumn{2}{c}{dev4}       &       \multicolumn{2}{c}{dev5}       &       \multicolumn{2}{c}{dev6}       \\
          std0 &       $d = -0.81$ & $p = 0.02$       & $d =  0.46$ & $p = 0.28$             & $d = +2.4$  & $p = 2\times10^{-6}$   \\
          std2 &        \multicolumn{2}{c}{  }        & $d = -3.2$  & $p = 5\times10^{-7}$   & $d = +0.26$ & $p = 0.096$            \\
          dev4 &        \multicolumn{2}{c}{  }        & $d = -1.4$  & $p = 2\times10^{-4}$   & $d = -3.3$  & $p = 2\times10^{-7}$   \\
          dev5 &        \multicolumn{2}{c}{  }        &        \multicolumn{2}{c}{  }        & $d = -2.1$  & $p = 4\times10^{-6}$   \\
          \\
               &                                        \multicolumn{6}{c}{\textbf{right IC}}                                       \\
          \hline
               &       \multicolumn{2}{c}{dev4}       &       \multicolumn{2}{c}{dev5}       &       \multicolumn{2}{c}{dev6}       \\
          std0 &       $d = -1.2$ & $p = 0.001$       &       $d = 0.06$ & $p = 0.84$        & $d = +2.0$  & $p = 2\times10^{-5}$   \\
          std2 &        \multicolumn{2}{c}{  }        & $d = -3.6$ & $p = 2\times10^{-7}$    & $d = -0.2$  & $p = 0.8$              \\
          dev4 &        \multicolumn{2}{c}{  }        &       $d = -1.3$ & $p = 0.0005$      & $d = -2.9$  & $p = 6\times10^{-7}$   \\
          dev5 &        \multicolumn{2}{c}{  }        &        \multicolumn{2}{c}{  }        & $d = -1.9$  & $p = 2\times10^{-5}$   \\
          \\
               &                                      \multicolumn{6}{c}{\textbf{left MGB}}                                         \\
          \hline
               &       \multicolumn{2}{c}{dev4}       &       \multicolumn{2}{c}{dev5}       &       \multicolumn{2}{c}{dev6}       \\
          std0 &  $d = -1.5$ & $p = 1.5\times10^{-4}$ &       $d = -0.37$ & $p = 0.29$       & $d = +1.7$  & $p = 8\times10^{-5}$   \\
          std2 &        \multicolumn{2}{c}{  }        & $d = -3.9$  & $p = 2\times10^{-7}$   & $d = -0.56$ & $p = 0.07$             \\
          dev4 &        \multicolumn{2}{c}{  }        &       $d = -1.2$ & $p = 0.0013$      & $d = -2.9$  & $p = 2\times10^{-7}$   \\
          dev5 &        \multicolumn{2}{c}{  }        &        \multicolumn{2}{c}{  }        & $d = -2.1$  & $p = 5\times10^{-6}$   \\
          \\
               &                                     \multicolumn{6}{c}{\textbf{right MGB}}                                         \\
          \hline
               &       \multicolumn{2}{c}{dev4}       &       \multicolumn{2}{c}{dev5}       &       \multicolumn{2}{c}{dev6}       \\
          std0 &       $d = -0.93$ & $p = 0.0066$     & $d = -0.16$ & $p = 1$                & $d = +1.2$  & $p = 9\times10^{-4}$   \\
          std2 &        \multicolumn{2}{c}{  }        & $d = -3.2$  & $p = 2\times10^{-7}$   & $d = -0.74$ & $p = 0.047$            \\
          dev4 &        \multicolumn{2}{c}{  }        & $d = -0.92$ & $p = 0.0086$           & $d = -2.3$  & $p = 2\times10^{-6}$   \\
          dev5 &        \multicolumn{2}{c}{  }        &        \multicolumn{2}{c}{  }        & $d = -1.6$  & $p = 1\times10^{-4}$   \\
          \hline
        \end{tabularx}
      \caption{\textbf{Statistics of the response differences between conditions.} Effect size is expressed as Cohen's $d$. Statistical significance was evaluated with two-tailed Ranksum tests between the distributions of the mean response in each ROI across subjects ($N = 19$). All $p$-values in the table are Holm-Bonferroni corrected for $4\times8 = 32$ comparisons. 
      \label{tab:effectsize}}
    \end{table*}

    To test the predictions of h2 (Figure~\ref{fig:design}C) we computed the correlation between the estimated BOLD response elicited by the different deviant positions in each SSA ROIs of the ICs and MGBs and the probability of finding the deviant in the $n$th position after hearing $n-1$ standards (namely: 1/3, 1/2 and 1, for deviant positions 4, 5, and 6, respectively; Figure~\ref{fig:rhos}). In agreement with the predictive coding hypothesis, we found a strong negative Pearson's correlation between predictability and BOLD responses in all four ROIs (left IC: $r = -0.72$, right IC: $r = -0.67$, left MGB: $r = -0.75$, right MGB: $r = -0.69$; $N = 19$ and $p < 10^{-37}$ in each of the four ROIs). These results showed that both IC and MGB display responses to the three deviant positions that are in accordance with the predictive coding hypothesis (h2, Figure~\ref{fig:design}C).

  \section*{Correlation of the responses with predictability at the single-subject level}

\begin{figure*}[t]
      \centering
      \includegraphics[width=\columnwidth]{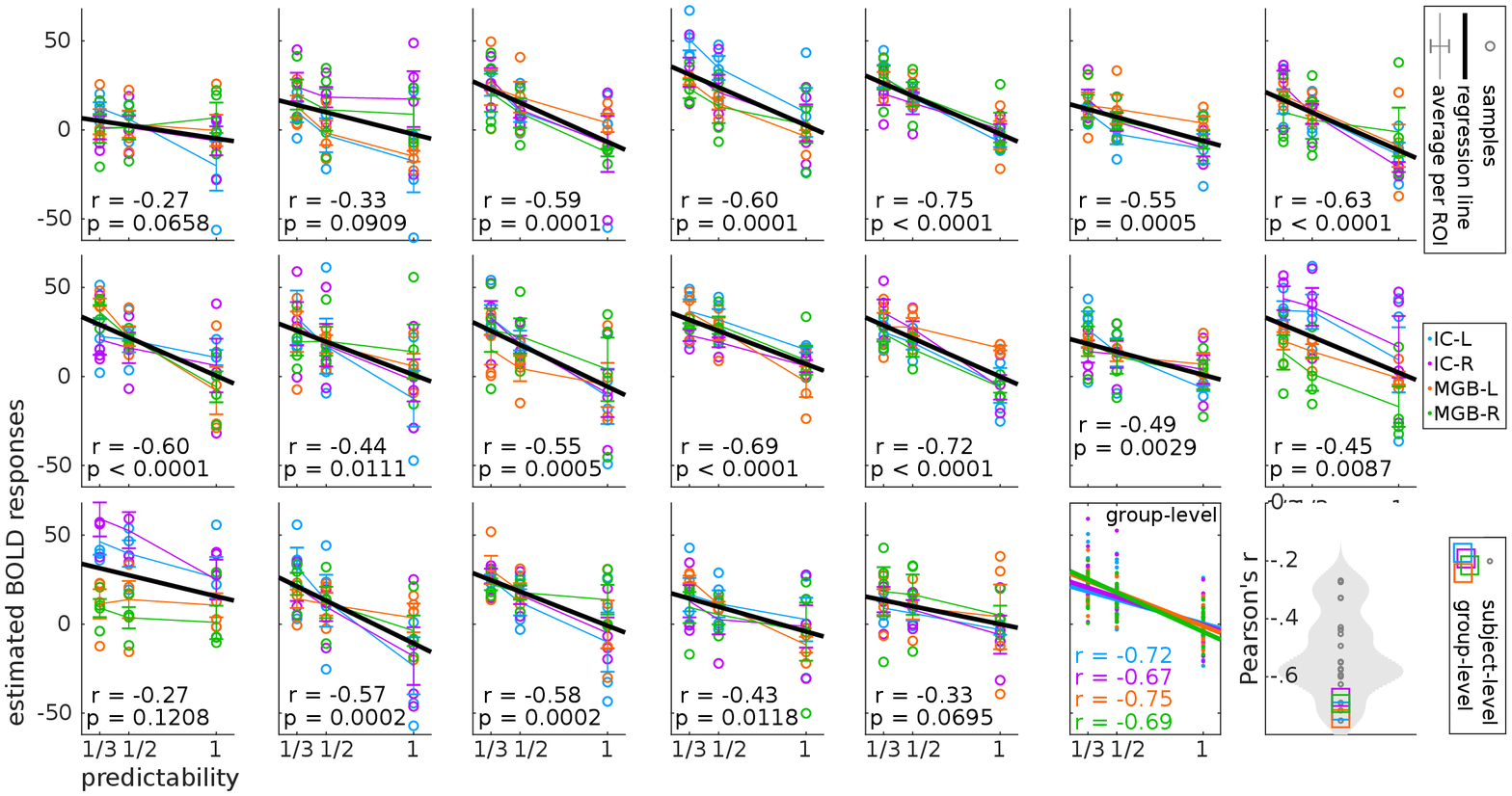}
      \caption{\textbf{Correlation between the estimated BOLD responses and predictability of the deviants.} Each plot displays, for each subject, the estimated BOLD response to the different deviant positions across voxels in the SSA ROIs in comparison to the predictability of the deviant. Error bars show the average and standard error per ROI (i.e., MGBs and ICs); black thick bars show the regression line across all samples from each subject. The last two plots show: 1) correlation per ROI at the group-level, and 2) kernel-density estimation of the distribution of correlation coefficients across subjects.
      \label{fig:rhos}}
    \end{figure*}

    To explore the robustness of the findings, we tested the correlation between the mean BOLD responses and deviant predictability at the single-subject level. We found negative correlations for each subject, with Pearson's $r$ ranging from $r = -0.27$ to $r = -0.72$, Figure~\ref{fig:rhos}. The correlations were statistically significant for 15 of the 19 subjects ($\mbox{p = 0.12}, \mbox{p = 0.09}, \mbox{p = 0.069}, p = 0.066$ for the non-significant correlations, and $p \in [0.012, 10^{-37}]$ for the significant ones; all $p$-values corrected for 19 comparisons; Pearson’s test comprised \mbox{$N = 4 \times 4 \times 3 = 48$} samples, corresponding to one sample for each ROI, run, and condition). 

  \section*{SSA can be abolished by making the deviant predictable}

    The result so-far suggested that the mesoscopic responses in the IC and MGB to the deviants can be interpreted as prediction error. If that is indeed the case, we expect that the deviant in position 6 would elicit similar responses as the standards after a deviant ($std2$), because the expectation of occurrence is the same (i.e., $P = 1$). In contrast, responses to a deviant in position 4 should show similar behaviour as deviants in traditional SSA designs; namely, higher response to the deviant than to the first standard ($std0$; \emph{deviant detection}) \cite{Cacciaglia2015, Gao2014, Malmierca2009}. The present data are consistent with both predictions: Response magnitudes for $dev6$ and $std2$ are similar (except for MGB-R) and the response to $dev4$ is significantly higher than to $std0$ in all four ROIs (Figure~\ref{fig:betas}; Cohen's $d < -0.8$, $p < 0.02$ corrected for multiple comparisons; Table~\ref{tab:effectsize}).

    The negligible differences between the responses to the expected deviant ($dev6$) and the standards after the deviant ($std2$) fit perfectly in the predictive coding framework: although the deviant is different from the standards in terms of frequency, it elicits the same response as a standard. Thus, SSA can be virtually abolished at the mesoscopic level by manipulating subjects' expectations; i.e., by rendering the deviant predictable.

  \section*{Predictive coding underlies the responses in both primary and secondary auditory pathways}

\begin{figure}[h]
      \centering
      \includegraphics[width=0.75\columnwidth]{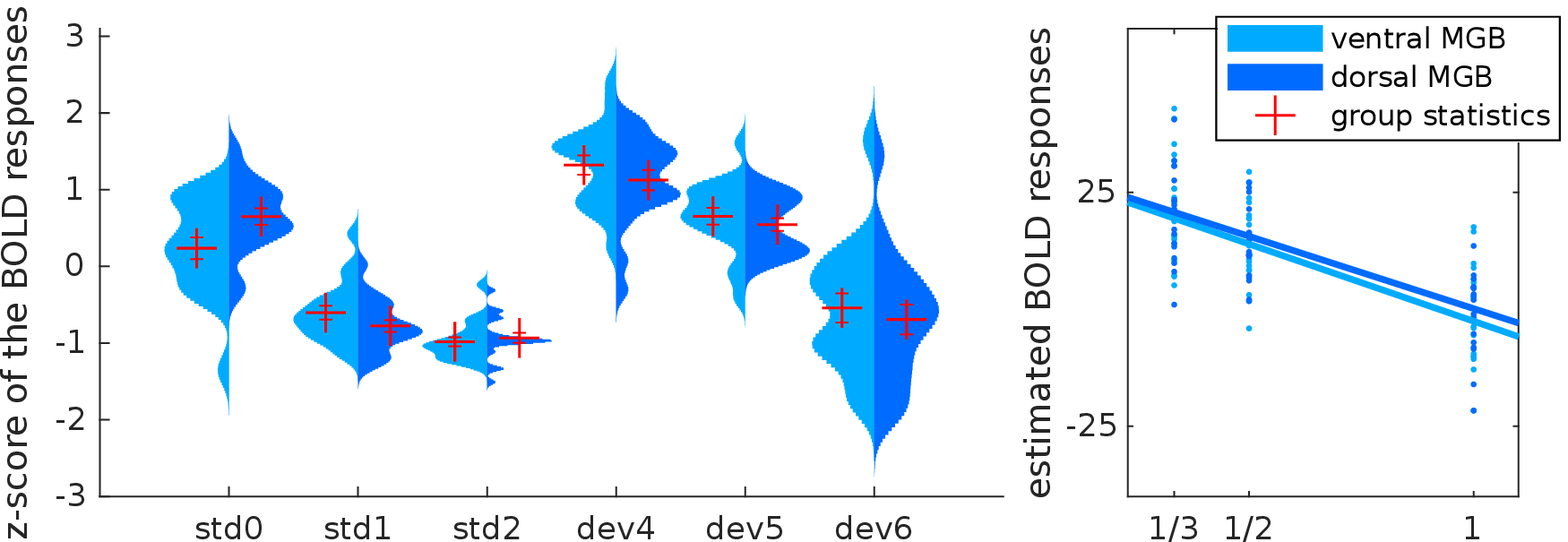}
      \caption{\textbf{Estimated BOLD responses in the subdivisions of the left MGB.} The figure shows a reproduction of the results shown in Figures~\ref{fig:betas} and~\ref{fig:rhos} considering the ROIs for the subdivisions of the left MGB reported in~\cite{Mihai2019}. The two subregions showed very similar response properties.
      \label{fig:subdivisions}}
    \end{figure}

    Last, we tested whether our results were specific to the primary (lemniscal) or the secondary (non-lemniscal) sections of the auditory pathway. Whilst the primary pathway is characterised by neurons that carry auditory information with high fidelity, the secondary pathway typically shows contextual and crossmodal effects \cite{Hu2003}. Although both the MGB and the IC contain subregions which contain either primary and secondary pathway components, distinguishing between the primary and secondary subsection of the IC and MGB non-invasively is technically challenging. A recent study \cite{Mihai2019} distinguished the primary subdivision of the left MGB (i.e., left ventral MGB) based on its location and functional properties from dorsal subsections of the MGB. We repeated the analyses above separately in each of these two parcellations to test if either mesoscopic SSA or the correlation of the elicited responses with predictability was specific to the secondary or the primary pathway. We found similar estimated BOLD responses (ventral vs. dorsal MGB; $dev4$: $d = 0.35$, $p = 0.43$; $dev5$: $d = 0.25$, $p = 0.35$; $dev6$: $d = 0.18$, $p = 0.56$) and similarly strong effect sizes in the dorsal ($r = -0.73$, $[-0.87, -0.47]$ 90\% confidence interval; $p < 10^{-37}$) and ventral ($r = -0.67$, $[-0.86, -0.31]$ 90\% confidence interval; $p < 10^{-37}$) subdivisions of the left MGB (Figure~\ref{fig:subdivisions}).

  \section*{Discussion}

    The main conclusion of our study is that processing in a primary sensory pathway can be appropriately described with a predictive coding framework. This conclusion is based on the finding that stimulus specific adaptation (SSA) in two auditory pathway stations (i.e., the MGB and IC) can be significantly modulated and even abolished by manipulating subjects’ expectations using abstract rules. Such response properties are expected under a predictive coding model (Figure~\ref{fig:design}C, h2), but not by habituation only (Figure~\ref{fig:design}C, h1). 

    The present study focused on auditory sensory pathway nuclei. Adaptation at early stages of the sensory pathways has, however, also been reported in the visual \cite{Dhruv2014}, olfactory \cite{ Fletcher2003} and somatosensory \cite{Kenshalo1980} pathways. We hypothesise that predictive coding mechanism could also explain these phenomena in the non-auditory sensory modalities. The adapting properties of predictive coding serve to optimise the dynamic range of sensory systems \cite{Brenner2000} and to maximise information transmission in the neural code by reducing the responses to expected stimuli \cite{Fairhall2001} and to redundant portions of the incoming sensory signal \cite{Huang2011}. The use of predictive strategies for sensory coding in the subcortical pathway indicates that information transmission is optimised at all points of the processing hierarchy. 

    Given the importance of predictive coding on sensory processing, atypical subcortical predictive coding could potentially result in profound repercussion at the cognitive level. For instance, subjects with developmental dyslexia, a disorder that is characterised by difficulties with processing speech sounds and with adaption to stimulus regularities \cite{Perrachione2016, Ahissar2006, Chandrasekaran2009} show altered responses in auditory pathway nuclei \cite{Diaz2012, Chandrasekaran2009} and atypical cortico-thalamic pathways \cite{MullerAxt2017, Tschentscher2019}. Thus, understanding the responses of the subcortical sensory pathway at a mechanistic level could also have valuable applications in clinical contexts.

  \section*{Methods}


      This study was approved by the Ethics committee of the Medical Faculty of the University of Leipzig, Germany (ethics approval number 273/14-ff). All listeners provided written informed consent and received monetary compensation for their participation.

    \subsection*{Participants}

      Nineteen German native speakers (12 female), aged 24 to 34 years (mean 26.6), participated in the study. None of them reported a history of psychiatric or neurological disorders, hearing difficulties, or current use of psychoactive medications. Normal hearing abilities were confirmed with pure tone audiometry (250\,Hz to 8000\,Hz; Madsen Micromate 304, GN Otometrics, Denmark) with a threshold equal to or below 25\,dB SPL. Participants were also screened for dyslexia (rapid automatised naming test of letters, numbers, and objects \cite{Denckla1974}; German LGVT 6-12 test \cite{Schneider2007}) and autism (Autism Spectrum Quotient \cite{Baron2001}). All scores were within the neurotypical range (RAN: maximum of 3.5 errors and $RT = 30$\,seconds across the four categories; AQ: all participants under a score of 23, below the cut-off value of 32; LGVT scores: all subjects where performing in the normal range).

    \subsection*{Experimental paradigm}

      All sounds were 50\,ms long (including 5\,ms in/out ramps) pure tones of frequencies 1455\,Hz, 1500\,Hz or 1600\,Hz, corresponding to three local minima of the power spectrum of the noise produced by the MRI during the scanning. From those three tones we constructed 6 standard-deviant frequency combinations that were used the same number of times across the experiment.

      Each tone sequence consisted of 7 repetitions of the standard stimulus and a single event of the deviant stimulus. Stimuli were separated by 700\,ms inter-stimulus-intervals (ISI), amounting to a total duration of 5300\,ms per sequence. To choose the ISI, we run a pilot behavioural study where we measured the reaction time to deviants 4, 5, and 6 with different ISIs. We took the shortest possible ISI that allowed the subjects to predict the entirely expected deviant, as revealed by a significant behavioural benefit in the RT for a deviant located in position 6.

      In each trial of the fMRI experiment, subjects listened to one tone sequence and reported, \emph{as fast and accurately as possible} using a button box with three buttons, the location of the deviant (4, 5 or 6). The inter-trial-interval (ITI) was jittered so that deviants were separated by an average of 5 seconds, up to a maximum of 11 seconds, with a minimum ITI of 1500\,ms. We chose such ITI properties to maximise the efficiency of the response estimation of the deviants \cite{Friston1999} while keeping a sufficiently long ITI to ensure that the sequences belonging to separate trials were not confounded.

      The experiment consisted in 4 runs with the same task. Each run contained 6 blocks of 10 trials. The 10 trials in each block used one of the 6 possible combinations of pure tones, so that all the sequences within each block had the same standard and deviant. Thus, within a block only the location of the deviant was unknown, while the frequency of the deviant was known. The order of the blocks within the experiment was randomised. The location of the deviant was pseudorandomised across all trials in each run so that each deviant location happened exactly 20 times per run but an unknown amount of times per block. This constraint allowed us to keep the same a priori probability for all deviant locations in each block. In addition, there were 23 silent gaps of 5300\,ms duration (i.e., null events of the same duration as the tone sequences) randomly located in each run \cite{Friston1999}.

      Each run lasted around 10 minutes, depending on the reaction times of the participant. The runs were separated by breaks of a minimum of 1 minute, during which the subjects could rest. Fieldmaps and a whole-head EPI (see~\emph{Data acquisition}) were acquired between the second and third run. The first run was preceded by a \emph{practice run} of four randomly chosen trials to ensure the subjects had understood the task. We scanned during the practice run in order to allow the subjects to undertake the training with scanning noise.

      The functional localiser’s experimental paradigm consisted of 6 runs of a duration of around 8 minutes each. In each run, subjects listened to 84 different natural sounds of 1000\,ms duration, interleaved with repetitions of a previous sound (5\% of the trials) and periods of silence of the same duration (null events, 5\% of the trials). Subjects were instructed to report any repetition of the sounds to keep their attention focus on the stimuli. The design is described in more detail in \cite{Moerel2015, Mihai2019}. 

    \subsection*{Data acquisition}

      MRI data were acquired using a Siemens Magnetom 7\,T scanner (Siemens Healthineers, Erlangen, Germany). We used a 32-channel head coil (NOVA Medical Inc, Wilmington MA, USA) for the functional localiser and an 8-channel head coil (RAPID Biomedical, Rimpar, Germany) for the main experiment. We used different coils in the two paradigms because the higher quality headphones, that were necessary to ensure that the pure tones with the smaller frequency difference were distinguishable in the main experiment, did not fit in the 32-channel coil. The 32-channel coil provided better SNR at a higher spatial resolution to better distinguish the subcortical nuclei. The 8-channel coil was sufficient to distinguish the overall subcortical activation.
    
      Functional MRI data were acquired using echo planar imaging (EPI) sequences. We used a field of view (FoV) of 132\,mm$\times$132\,mm and partial coverage with 30 slices. This volume was oriented in parallel to the superior temporal gyrus such that the slices encompassed the IC, the MGB, and the superior temporal gyrus. In addition, we acquired three volumes of an additional whole-head EPI with the same parameters (including the FoV) and 80 slices during resting to aid the coregistration process (see~\emph{Data preprocessing}).

      The EPI sequences used for the main experiment had the following acquisition parameters: \mbox{TR = 1600\,ms}, TE = 19\,ms, flip angle 65$^{\circ}$, GRAPPA with acceleration factor 2 \cite{Griswold2002}, 33\% phase oversampling, matrix size $88\times88$, FoV 132\,mm$\times$132\,mm, phase partial Fourier 6/8, voxel size 1.5\,mm isotropic, interleaved acquisition, and anterior to posterior phase-encode direction. During functional MRI data acquisition, heart rate and respiration rate were acquired using a BIOPAC MP150 system (BIOPAC Systems Inc., Goleta, CA, USA). Data of the functional localiser was acquired using a sparse variation of this sequence with a higher spatial resolution (1.1\,mm isotropic), a 32 channel coil, and TR = 2800\,ms, TA = 1600\,ms. Sparse imaging was used to leave a 1200\,ms window of silence between the acquisitions to present the stimuli (for more details see~\cite{Moerel2015, Mihai2019}).

      Structural images were recorded using an MP2RAGE \cite{Marques2010} T1 protocol with 700 $\mu$m isotropic resolution, TE = 2.45\,ms, TR = 5000\,ms, TI1 = 900\,ms, TI2 = 2750\,ms, flip angle 1 = 5$^{\circ}$, flip angle 2 = 3$^{\circ}$, \mbox{FoV = 224\,mm$\times$224\,mm}, GRAPPA acceleration factor 2.

      Stimuli were presented using MATLAB (The Mathworks Inc., Natick, MA, USA) with the Psychophysics Toolbox extensions \cite{Brainard1997} and delivered through an MrConfon amplifier and headphones (MrConfon GmbH, Magdeburg, Germany). Loudness was adjusted independently for each subject before starting the data acquisition to a comfortable level.

    \subsection*{Data preprocessing}

      The preprocessing pipeline was coded in Nipype 1.1.2 \cite{Gorgolewski2011}, and carried out using tools of the Statistical Parametric Mapping toolbox, version 12 (SPM); Freesurfer, version 6 \cite{Fischl2002}; the FMRIB Software Library, version 5 (FSL) \cite{Jenkinson2012}); and the Advanced Normalization Tools, version 2.2.0 (ANTS) \cite{Avants2011}. All data were coregistered to the Montreal Neurological Institute (MNI) MNI152 1\,mm isotropic symmetric template. 
     
      First, we realigned the functional runs. We used SPM's \emph{FieldMap Toolbox} to calculate the geometric distortions caused in the EPI images due to field inhomogeneities. Next, we used SPM's  \emph{Realign and Unwarp} to perform motion and distortion correction on the functional data. Motion artefacts, recorded using SPM's ArtifactDetect, were later added to the design matrix (see~\emph{Estimation of the BOLD responses}).

      Next, we processed the structural data. We first masked the structural data to eliminate voxels that contain air, scalp, skull and cerebrospinal fluid. The masks were computed by first segmenting the white matter with SPM's \emph{Segment} and applied with \emph{FSLmaths}. Then, we used Freesurfer's recon-all routine to calculate the boundaries between grey and white matter (these are necessary to register the functional data to the structural images) and ANTs to compute the transformation between the structural images and the MNI152 symmetric template.

      Last, we coregistered the functional data to the MNI152 space. The transformation between the functional runs and the structural image was computed with using Freesurfer’s \emph{BBregister} using the boundaries between grey and white matter of the structural data and the whole-brain EPI as an intermediate step. The final functional-to-MNI transformation, computed as the concatenation of the functional-to-structural and structural-to-MNI transformations, was then applied using ANTs. Note that, since the resolution of the MNI space (1\,mm isotropic) was higher than the resolution of the functional data (1.1\, mm isotropic and 1.5\,mm isotropic), the transformation resulted in a spatial oversampling.

      All the preprocessing parameters, including the smoothing kernel size, were fixed before we started fitting the general linear model (GLM) and remained unchanged during the subsequent steps of the data analysis.

      Physiological (heart rate and respiration rate) data were processed by the PhysIO Toolbox \cite{Kasper2017}, that computes the Fourier expansion of each component along time and adds the coefficients as covariates of no interests in the model's design matrix. 

    \subsection*{Estimation of the BOLD responses}

      First level and second level analyses were coded in Nipype and carried out using SPM. Statistical analyses of the model estimations in the SSA ROIs were carried out using custom code in MATLAB. BOLD data acquired during the practice run was not included in the analysis.

      The coregistered data were first smoothed using a 2\,mm full-width half-maximum kernel Gaussian kernel with SPM's \emph{Smooth}. 

      In the main experiment, the first level GLM's design matrix included 6 conditions: first standard (std0), standards before the deviant (std1), standards after the deviant (std2), and deviants in locations 4, 5, and 6 (dev4, dev5, and dev6, respectively; Figure~\ref{fig:design}). Conditions std1 and std2 were modelled using linear parametric modulation \cite{ODoherty2007}, whose linear factors were coded according to the position of the sound within the sequence (see Supplementary Figure S1). We modelled the first standard separately from the remaining standards preceding the deviant so that we could perform a contrast comparing the responses to the first and the adapted standards to locate voxels showing adaptation. We modelled the standards preceding and following the deviant separately because we cannot propose a set of linear factors simultaneously valid for both, std1 and std2. On top of the main regressors, the design matrix also included the physiological PhysIO and artefact regressors of no-interest.

      In the first level analysis of the main experiment, we computed two $t$-test contrasts: 1) ``$\text{std0} - \text{std2} > 0$'' (adaptation), and 2) ``$\text{dev4} - \text{std2} > 0$'' (deviant detection). In the second level analysis, we used the same contrasts across subjects.

      In the functional localiser, the first level GLM's design matrix included just one condition, labelled as \emph{sound}, corresponding to the presentation of a sound, and the physiological PhysIO and artefact regressors of no-interest. In the first level analysis, we computed the $t$-test contrast ``$\text{sound} > 0$''. The same contrast was used in the second level analysis across subjects.

    \subsection*{Calculation of the IC and MGB ROIs}

      We started considering four prior regions for each of the four nuclei of interest based on the anatomical references of the MNI template. These prior regions were used to mask the t-maps of the contrast defined for the functional localiser using \emph{FSLmaths}. We computed the IC and MGB ROIs (yellow patches in Figure~\ref{fig:ROIs}) by thresholding the masked t-maps at increasing t-values until the ROI had the volume of the subcortical nuclei of interest (i.e., $146$ voxels for the ICs and $152$ voxels for the MGBs based on in-vivo 7T MRI \cite{Sitek2019}; thresholds were $t > 6.00$, $t > 5.57$, $t > 8.58$, and $t > 8.87$, for the right MGB, left MGB, right IC, and left IC, respectively).

      We used SPM to calculate the adaptation (Figure~\ref{fig:ROIs}, blue patches) and deviant detection (red patches) ROIs, defined as the sets of voxels within the IC and MGB ROIs that responded significantly to the contrasts $std0 > std2$ and $dev4 > std2$, respectively. Significance was defined as $p < 0.05$, Bonferroni-corrected for the number of voxels within each of the IC/MGB ROIs. The SSA ROIs (Figure~\ref{fig:ROIs}, purple patches) were then calculated as the intersection between the adaptation and deviant detection ROIs using \emph{FSLmaths}.

  \bibliography{bib}
  \bibliographystyle{ieeetr}

  \newpage

  \section*{Supplementary materials}

    \subsection*{Supplementary text S1} 

      In the main text, we have assumed for simplicity that the habituation hypothesis predicts the same response for the deviants independently of their locations. The validity of this assumption depends, however, on two factors: 1) the time-lapse from the previous deviant, and 2) the time constant of the habituation $\tau_h$. If $\tau_h$ is much shorter than the expected time lapse between deviants, then the hypothesis of the main text is correct. Otherwise, since deviants in earlier locations are more likely to occur within a shorter time-lapse to the previous deviant, on average, $dev4$ will show a weaker response than $dev5$, and $dev5$ would show a weaker response than $dev6$. Note that the direction of this potential effect is opposite to that of the predictive coding hypothesis.

      A third scenario, not included in the main text for the sake of conciseness, would be that the neural responses correspond to precision-weighted predictive coding, where the model used to perform the predictions is based on the local statistics of the stimuli only (i.e., ignoring the abstract rules). In this scenario, earlier deviants would violate expectations with less precision than later deviants, since there would have been less preceding standards contributing to the build-up of the internal model of the repeated stimulus. Since higher precisions elicit stronger levels of prediction error, this scenario also predicts stronger responses for later deviants.

      The agreement between precision-weighted predictive coding based on local statistics and habituation illustrate, once again, the confound between predictive coding and habituation ubiquitous to stimulus-specific adaptation. Only with the introduction of abstract rules can predictive coding be entirely disassociated from habituation.

    \begin{figure}[h]
      \centering
      \includegraphics[width=1\columnwidth]{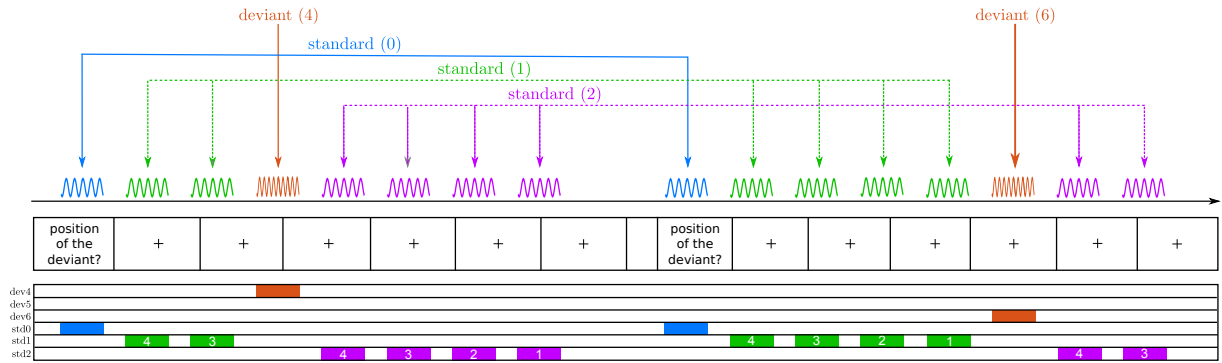}
      \caption{\textbf{Schematic of the GLM's design matrix.} An example of the GLM design matrix section corresponding to the regressors of interest prior to the convolution with the haemodynamic response function. The example includes two trials with two different deviant locations. The six regressors of interest were standard 0, standard 1, standard 2, deviant 4, deviant 5 (not shown), deviant 6. The standards were parametrically modulated and the modulation was equal to the inverted index of the standard within the sequence (i.e., 4 for the first standard 1/first standard 2, 3 for the second standard 1/second standard 2, etc; note that, since the modulation was mean-corrected before the fitting of the GLM, the absolute values of the modulation are not relevant). 
      \label{fig:s1}}
    \end{figure}
\end{document}